\documentclass[aps,prl,twocolumn,floatfix,superscriptaddress]{revtex4}

\usepackage{graphicx}
\usepackage{amsmath,verbatim}
\usepackage{color}
\usepackage{bm}% bold math

\input vmc.def

\begin{document}

\title{Alleviation of the Fermion-sign problem by optimization of many-body wave functions}

\author{C. J. Umrigar}
\author{Julien Toulouse}
\affiliation{Theory Center and Laboratory of Atomic and Solid State Physics, Cornell University, Ithaca, NY 14853, USA}

\author{Claudia Filippi}
\affiliation{Instituut Lorentz, Universiteit Leiden, Niels Bohrweg 2, Leiden, NL-2333 CA, The Netherlands}

\author{S. Sorella}
\affiliation{INFM-Democritos National Simulation Centre, and SISSA, Trieste, Italy}

\author{R. G. Hennig}
\affiliation{Materials Science and Engineering, Cornell University, Ithaca, NY 14853, USA}

\begin{abstract}
We present a simple, robust and highly efficient method for optimizing all parameters
of many-body wave functions in quantum Monte Carlo calculations, applicable to
continuum systems and lattice models.
Based on a strong zero-variance principle, diagonalization of the Hamiltonian matrix in the space spanned by the wave function and its derivatives determines the optimal parameters.
It systematically reduces the fixed-node error, as demonstrated by the calculation of the
binding energy of the small but challenging C$_2$ molecule to the experimental accuracy of 0.02 eV.

\end{abstract}

\date{\today}
\maketitle

Many important problems in computational physics and chemistry can be reduced to the computation
of the dominant eigenvalues of matrices or integral kernels.
For problems with many degrees of freedom, Monte Carlo approaches such as
diffusion, reptation
and transfer-matrix Monte Carlo are among the most accurate and efficient
methods~\cite{ProjectorMC}.
Their efficiency and in most cases accuracy rely crucially on
good approximate eigenstates.
For Fermionic systems, the antisymmetry constraint leads to the Fermion-sign problem that forces practical
implementations to fix the nodes to those of an approximate trial wave function.
The resulting fixed-node error is the main uncontrolled error in quantum Monte Carlo (QMC).
Despite great effort aimed at developing algorithms that eliminate
this error entirely~\cite{FermionSign}, to date there is no generally applicable solution, therefore
systematic improvement of the wave function by optimization of an increasing number of
variational parameters is the most practical approach for reducing this error.

In this letter we develop a robust and highly efficient method for optimizing
all parameters in approximate wave functions.
Application to the small but challenging C$_2$ molecule demonstrates that the
method systematically eliminates the fixed-node error.
The method is based on energy minimization in variational Monte Carlo (VMC) and extends
the zero-variance linear optimization method~\cite{NightingaleAlav01}
to nonlinear parameters. The approach is simpler than the Newton method~\cite{UmrigarFilippi05} as it
does not require second derivatives of the wave function.
Moreover, in contrast to the perturbative optimization method~\cite{ScemamaFilippi06},
it enables the simultaneous parameter optimization of both the Jastrow and the determinantal parts
of the wave function with equal efficiency.

\begin{comment}
Here, we develop a robust and highly efficient method for optimizing all the parameters
of many-body wave functions based on energy minimization in variational Monte Carlo (VMC), which extends
the zero-variance linear optimization method of Nightingale and Melik-Alaverdian~\cite{NightingaleAlav01}
to nonlinear parameters. The approach is simpler than the Newton method~\cite{UmrigarFilippi05} as it
does not require second derivatives of the wave function with respect
to the parameters. Moreover, in contrast to the perturbative optimization method~\cite{ScemamaFilippi06},
it allows us to simultaneously optimize the parameters in both the Jastrow and the determinantal parts of the wave function with equal efficiency~\cite{ToulouseUmrigar06}.
We illustrate the effectiveness of the method on the small but challenging C$_2$ molecule,
achieving a systematic improvement of the DMC energy as the number of Slater determinants
increases and reaching chemical accuracy for the equilibrium well depth.
\end{comment}

\vskip 2mm \noindent {\it Form of wave functions.} \hskip 2mm
We employ %Jastrow-Slater
$N$-electron wave functions which depend on $\Nopt$ variational parameters collectively denoted by
$\pvec = (\cvec, \alphavec, \lambdavec)$ and $3N$ electronic coordinates, $\Rvec$,
\beq
\Psi(\pvec,\Rvec) = J(\alphavec,\Rvec) \sum_{i=1}^{\NCSF} c_i C_i(\lambdavec,\Rvec),
\eeq
where $J(\alphavec,\Rvec)$ is a Jastrow factor that contains electron-nuclear, electron-electron and electron-electron-nuclear
terms.  Each of the $\NCSF$ configuration state functions (CSFs), $C_i(\lambdavec,\Rvec)$, is a symmetry-adapted linear combination of Slater determinants
of single-particle orbitals $\phi_\mu(\rvec)$ which are themselves
expanded in terms of $N_{\rm basis}$ basis functions $\chi_\nu(\rvec)$:
$\phi_\mu(\rvec) = \sum_{\nu=1}^{N_{\rm basis}} \lambda_{\mu \nu} \chi_\nu(\rvec)$.
The variational parameters are the linear CSF coefficients $\cvec$,
and the nonlinear Jastrow parameters $\alphavec$
and expansion coefficients of the orbitals $\lambdavec$. In practice, we optimize only
$\NCSF -1$ CSF coefficients (as the normalization of the wave function is irrelevant)
and a set of non-redundant orbital rotation parameters~\cite{SchautzFilippi04,ToulouseUmrigar06}.

\vskip 2mm \noindent {\it Optimization of wave functions.} \hskip 2mm
We extend the zero-variance method of Nightingale and Alaverdian~\cite{NightingaleAlav01} for linear parameters
to nonlinear parameters.
At each optimization step, the wave function is expanded to linear-order in
$\Delta \pvec = \pvec \!-\! \pvec^0$ around the current parameters $\pvec^0$,
\beq
\Psi_{\text{lin}}(\pvec,\Rvec) = \Psi_0(\Rvec) +  \sumilNopt \Delta p_i \, \Psi_i(\Rvec),
\label{linear_approx}
\eeq
where $\Psi_0 = \Psi(\pvec_0)$ is the current wave function and
$\Psi_i=\left(\partial \Psi(\pvec)/\partial p_i \right)_{\pvec=\pvec_0}$
are the $\Nopt$ derivatives of the wave function with respect to the parameters.
On an \textit{infinite} Monte Carlo (MC) sample, the % optimal
parameter variations $\Delta \pvec$
minimizing the energy calculated with the linearized wave function of Eq.~(\ref{linear_approx})
are the lowest eigenvalue solution of the generalized eigenvalue equation
\beq
\Hvec \Delta \pvec = E \, \Svec \Delta \pvec,
\label{eigenvalue_eq}
\eeq
where $\Hvec$ and $\Svec$ are the Hamiltonian and overlap matrices in the
($\Nopt+1$)-dimensional basis formed by the current wave function and its derivatives
$\left\{ \Psi_0, \Psi_1, \Psi_2, \cdots, \Psi_{\Nopt} \right\}$ and $\Delta p_0 = 1$.
On a \textit{finite} MC sample, following Ref.~\onlinecite{NightingaleAlav01},
we estimate these matrices by
\beq
H_{ij} = \left\langle {\Psi_i \over \Psi_0} {H \Psi_j \over \Psi_0}\right\rangle_{\Psi_0^2}, \;\;\;\;\;
S_{ij} = \left\langle {\Psi_i \over \Psi_0} {\Psi_j \over \Psi_0}\right\rangle_{\Psi_0^2},
\label{S_and_H}
\eeq
where $H$ is the electronic Hamiltonian.
We employ the notation that
$\Langle{\Psi_i \over \Psi_0}{A \Psi_j \over \Psi_0}\Rangle_{\Psi_0^2}$
denotes the statistical estimate of
$\brakett{\Psi_i}{\hat{A}}{\Psi_j} / \braket{\Psi_0}{\Psi_0}$ evaluated using MC configurations sampled from $\Psi_0^2$.
The estimator for the matrix ${\bf H}$ in Eq.~(\ref{S_and_H}) is \textit{nonsymmetric} and is \textit{not}
the symmetric Hamiltonian matrix that one would obtain by minimizing the energy of the finite MC sample.
In fact, as shown in Ref.~\onlinecite{NightingaleAlav01}, it is only this nonsymmetric estimator
for ${\bf H}$ that leads to a strong zero-variance principle: the variance of the parameter changes
$\Delta \pvec$ in Eq.~(\ref{eigenvalue_eq}) vanishes not only in the limit that $\Psi_0$ is the
exact eigenstate but even in the limit that
$\Psilin$ with optimized parameters is an exact eigenstate.
In practice, for the wave functions we employ, and optimizing on small MC samples,
the asymmetric Hamiltonian results in 1 to 2 orders of magnitude smaller fluctuations
in the parameter values.

Quite generally, methods that minimize the energy of a finite MC sample are stable only if a
much larger number of MC configurations is employed than is necessary for the variance minimization
method~\cite{UWW88} because the energy of a finite sample is not bounded from
below but the variance is.  However, it is possible to devise simple modifications of these energy-minimization
methods
that require orders of magnitude fewer MC configurations.
Both the zero-variance linear method employed here and
the modified Newton method~\cite{UmrigarFilippi05} are examples of such modifications.

Having obtained the parameter variations $\Delta \pvec$ by solving Eq.~(\ref{eigenvalue_eq}),
there is no unique way to update the parameters in the wave function.
The simplest procedure of incrementing the current parameters by
$\Delta \pvec$, $\pvec_0 \to \pvec_0 + \Delta \pvec$ works for the
linear parameters but is not guaranteed to work for the nonlinear parameters if
the linear approximation of Eq.~(\ref{linear_approx}) is not good.
A better but more complicated procedure is to fit the original wave function form to
the optimal linear combination.  We next discuss a simple prescription that avoids
doing this fit.

At first, it may appear that %, for the chosen functional form of $\Psi$,
nothing can be done to make the linear approximation better, but this is
in fact not the case.
One can exploit the freedom of the normalization of the wave function to
alter the dependence on the nonlinear parameters as follows.
Consider the differently-normalized wave function
$\Psib(\pvec,\Rvec) = N(\pvec) \Psi(\pvec,\Rvec)$ such that
$\Psib(\pvec_0,\Rvec) = \Psi(\pvec_0,\Rvec) \equiv \Psi_0(\Rvec)$ %for the current value of parameters,
and $N(\pvec)$ depends only on the nonlinear parameters.
Then, the derivatives of $\Psib(\pvec)$ are
\beq
\Psib_i = \Psii + N_i \Psi_0, \hbox{\hskip 3mm where \hskip 3mm}
N_i=\left(\partial N(\pvec)/\partial p_i \right)_{\pvec=\pvec_0},
\eeq
with $N_i=0$ for linear parameters.
The first-order expansion of $\Psib(\pvec)$ after optimization is
\beq
\Psib_{\text{lin}} = \Psi_0 + \sumilNopt \Delta \pbar_i \, \Psib_i .
\label{linear_approx2}
\eeq
Since $\Psib_{\text{lin}}$ and $\Psi_{\text{lin}}$ were obtained by optimization in the same variational space
they must be proportional to each other,
so $\Delta \pbarvec$ are related to $\Delta \pvec$ by a uniform rescaling
\beq
\Delta \pbarvec &=& { \Delta \pvec \over 1 - \sumilNopt N_i \Delta p_i}.
\label{orig_to_semiorthog}
\eeq
Since the rescaling factor can be anywhere between $-\infty$ and $\infty$
the choice of normalization can affect not only the magnitude of the parameter changes but even the sign.

For the nonlinear parameters, we have found that, in all cases considered here, a fast and
stable optimization is achieved if $N_i$ are determined by imposing the
condition that each derivative $\Psib_i$ is orthogonal to a linear
combination of $\Psi_0$ and $\Psi_{\text{lin}}$, i.e.,
$\braket{\xi {\Psi_0 \over \vert\vert \Psi_0 \vert\vert}
 + (1-\xi) {\Psi_{\text{lin}} \over \vert\vert \Psi_{\text{lin}} \vert\vert}}{\Psib_i} =0$,
where $\xi$ is a constant between 0 and 1,
resulting in %\begin{equation}
\beq
N_i &=& -{\xi D S_{0i} + (1-\xi)(S_{0i}+ \sumjnonlin S_{ij} \Delta p_j) \over \xi D + (1-\xi)(1+\sumjnonlin S_{0j} \Delta p_j)},
\label{N_i}
\eeq
with $D=\sqrt{1+2\sumjnonlin S_{0j} \Delta p_j + \sumjknonlin S_{jk} \Delta p_j \Delta p_k}$,
where the sums are over only the nonlinear parameters.
The simple choice $\xi=1$ first used by Sorella in the context of the stochastic reconfiguration
method~\cite{Sorella01}
leads in many cases to good parameter variations, but in some cases can result in arbitrarily large parameter
variations that may or may not be desirable.
The safer choice $\xi=0$ minimizes the norm of the linear wave function change
$\vert\vert \Psib_{\text{lin}} - \Psi_0 \vert\vert$
in the case that only the nonlinear parameters are varied,
but, it can yield arbitrarily small parameter changes even far from the energy minimum.
In contrast, the choice $\xi=1/2$, imposing
$\vert\vert \Psib_{\text{lin}} \vert\vert = \vert\vert \Psi_0 \vert\vert$,
guarantees finite parameter changes, until the energy minimum is reached.

If instead of finding the parameter changes from Eqs.~(\ref{eigenvalue_eq}) and (\ref{orig_to_semiorthog})
one expands the Rayleigh quotient
$ \brakett{\Psiblin}{\Hhat}{\Psiblin} / \braket{\Psiblin}{\Psiblin}$ with $\xi=1$ to second order in $\Delta \pbarvec$
one recovers the stochastic reconfiguration with Hessian acceleration (SRH) method with $\beta=0$ of Ref.~\onlinecite{Sorella05}.
It turns out that the SRH method is much less stable and converges more slowly, particularly for large systems with many parameters.

\vskip 2mm \noindent {\it Stabilization of the optimization.} \hskip 2mm
When the parameter values are far from optimal or if the MC sample used to
evaluate the ${\bf H}$ and ${\bf S}$ matrices is very small, then the updated parameters
may be worse than the original ones. However, it is possible to devise a scheme to stabilize
the method in a manner similar to that used for the modified Newton method~\cite{UmrigarFilippi05}.
Stabilization is achieved by adding a positive constant, $\adiag \geq 0$, to the diagonal of
the Hamiltonian matrix except for the first element:
$H_{ij} \to H_{ij} + \adiag \delta_{ij} (1 - \delta_{i0})$.
As $\adiag$ becomes larger the parameter variations $\Delta \pvec$ become smaller and rotate
towards the steepest descent direction.
In practice, the value of $\adiag$ is automatically adjusted at each optimization step.
Once the matrices $\Hvec$ and $\Svec$ have been computed,
three values of $\adiag$ differing from each other by factors of 10 are used
to predict three new wave functions. A short MC run is then performed using correlated
sampling to compute energy differences of these wave functions more accurately than the energy
itself. The optimal value of $\adiag$ is then calculated by parabolic interpolation on these three energies,
with some bounds imposed.

\vskip 2mm \noindent {\it Results.} \hskip 2mm
We demonstrate the performance of our optimization method on the $^1\Sigma_g^+$ ground state of the C$_2$
molecule at the experimental equilibrium interatomic distance of $2.3481$ Bohr, employing a
scalar-relativistic Hartree-Fock (HF) pseudopotential~\cite{BurkatzkiFilippiDolg06} with a large Gaussian
polarized valence quintuple-zeta one-electron basis
(12 $s$, 10 $p$ and 4 $d$ functions contracted to 5 $s$, 5 $p$ and 4 $d$ functions)
to ensure basis-set convergence.
The quantum chemistry package GAMESS~\cite{gamess} is used to obtain
multi-configurational self-consistent field (MCSCF) wave functions in complete active spaces (CAS)
generated by distributing $n$ valence electrons in $m$ orbitals [CAS($n$,$m$)].
We also employ restricted active space RAS($n$,$m$) wave functions consisting of a truncation of
the CAS($n$,$m$) wave functions to quadruple excitations.
By applying a variable cutoff on the CSF coefficients, only the Slater determinants with the largest CSF coefficients
are retained in these wave functions, which are then multiplied by a Jastrow factor with essentially all free
parameters chosen to be zero to form our starting trial wave functions.

\begin{figure}
\includegraphics[width=8cm]{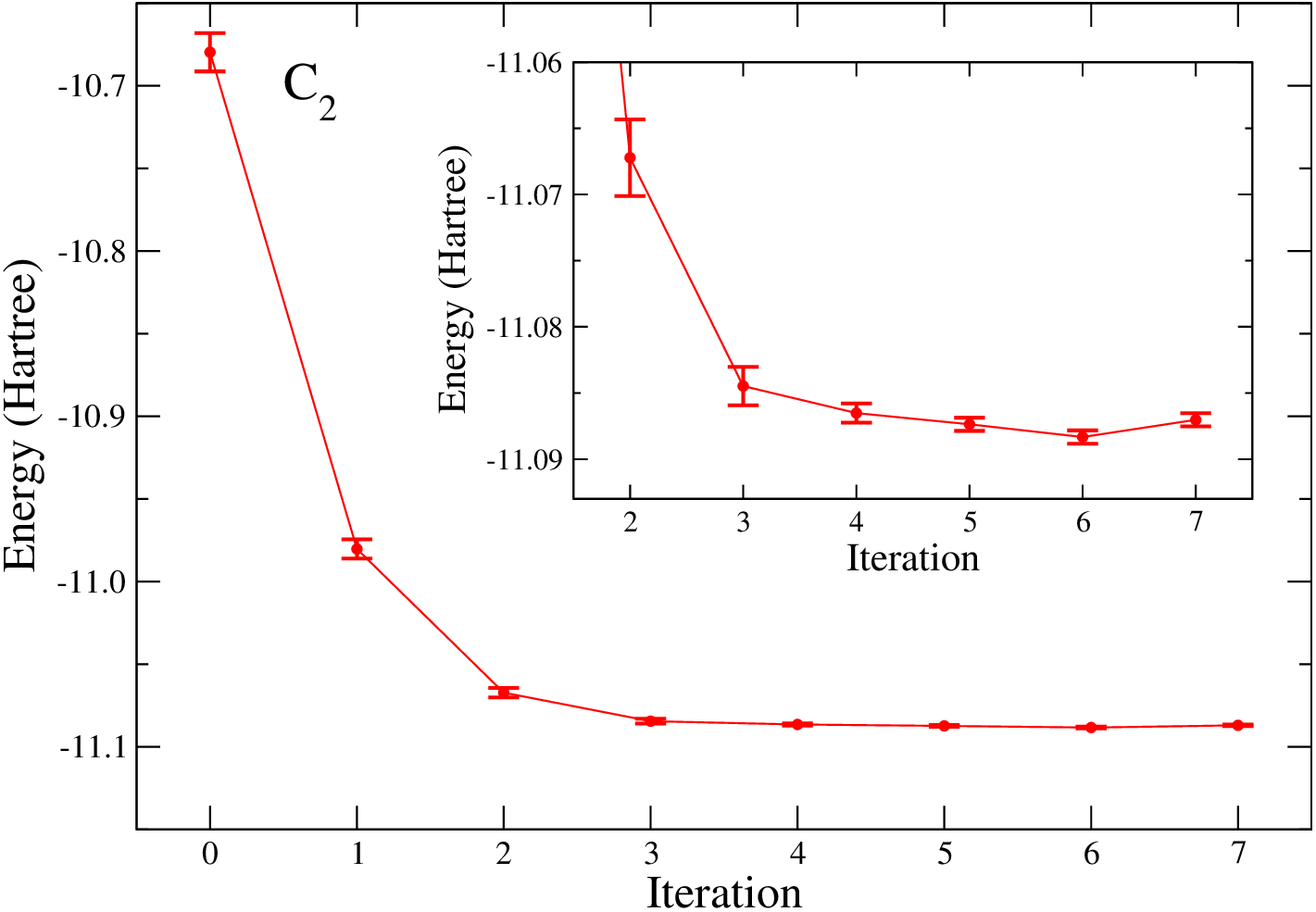}
\caption{
Convergence of the VMC total energy of the C$_2$ molecule with a
HF pseudopotential~\cite{BurkatzkiFilippiDolg06} when simultaneously optimizing 24 Jastrow,
73 CSF and 174 orbital
parameters for a truncated CAS(8,14) wave function.
The insert is an enlargement of the last iterations.}
\label{energy_c2}
\vspace{-0.4cm}
\end{figure}

Fig.~\ref{energy_c2} illustrates the convergence of the VMC energy during the simultaneous optimization of 24 Jastrow,
73 CSF and 174 orbital
parameters for a truncated CAS(8,14) wave function.
The energy
converges to an accuracy of about $10^{-3}$
Hartree in 5 iterations, making it the most rapidly convergent method for optimizing all the parameters
in Jastrow-Slater wave functions.

\begin{figure}
\includegraphics[width=8cm]{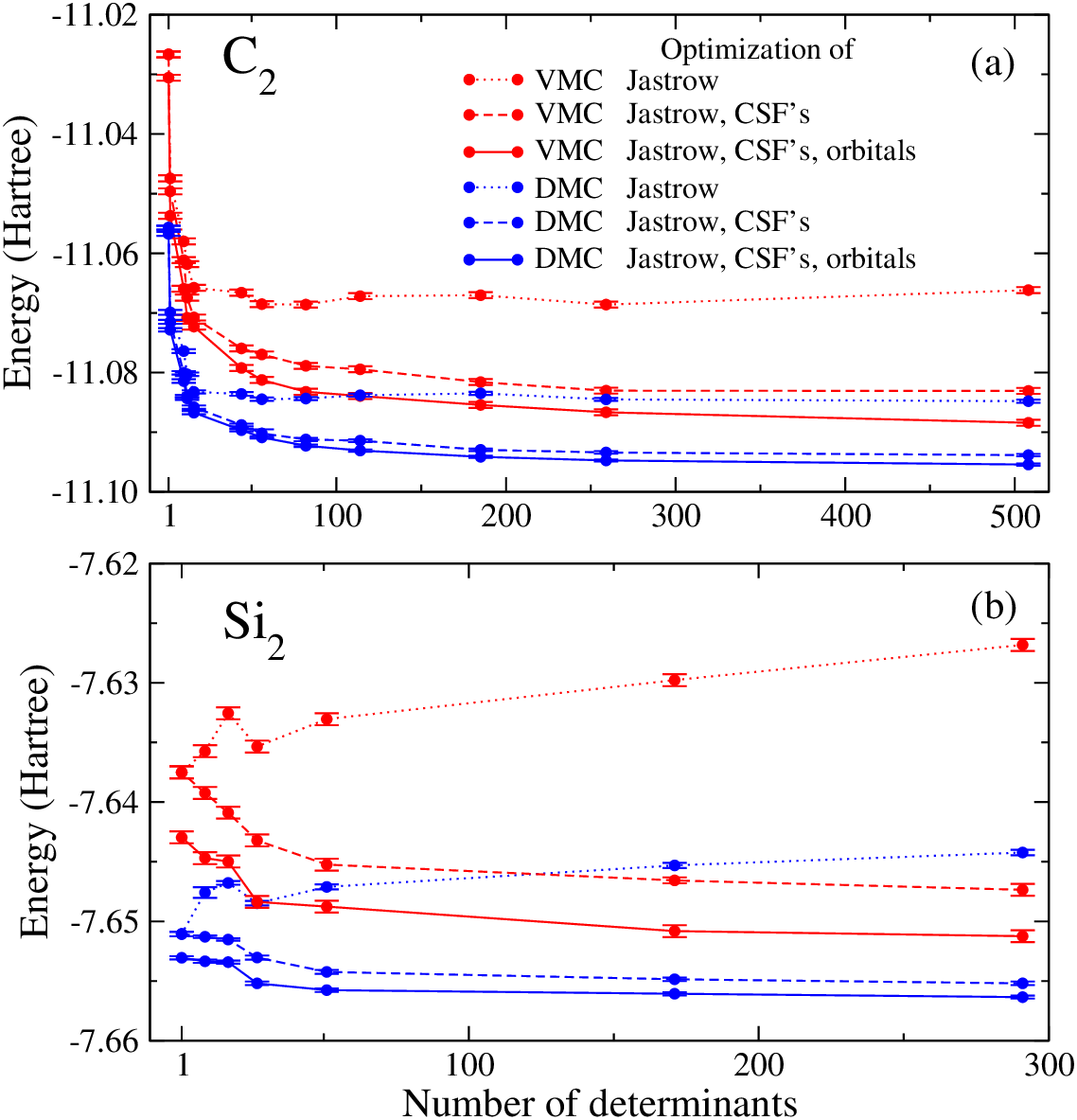}
\caption{(Color online) VMC and DMC total energies of the C$_2$ and Si$_2$ molecules,
versus the number of determinants
in truncated Jastrow-Slater CAS(8,14) wave functions,
for different levels of optimization.
Both VMC and DMC energies decrease monotonically if the CSF coefficients are reoptimized in VMC
but not if only the Jastrow factor is optimized. %the CSF coefficients are kept fixed at their MCSCF values.
}
\label{c2}
\vspace{-0.4cm}
\end{figure}

Fig.~\ref{c2} shows the total VMC and DMC energies of the C$_2$ molecule as a function of the
number of determinants retained in truncated Jastrow-Slater CAS(8,14) wave functions.
For comparison, the convergence of the $^3\Sigma_g^+$ ground state energy of Si$_2$ is also shown.
For each of VMC and DMC, there are three curves.
For the upper curve, only the Jastrow parameters are optimized and the CSF and orbital coefficients
are fixed at their MCSCF values.  For the middle curve, the Jastrow and CSF parameters
are optimized simultaneously while, for the lower curve, the Jastrow, CSF and orbital parameters
are optimized simultaneously.
With fixed CSF coefficients, the energy does not improve monotonically with the number of determinants.
In contrast, if the CSF coefficients are reoptimized then of course the VMC energy improves
monotonically.  Remarkably, so does the DMC energy, implying that the nodal hypersurface
of the wave function is monotonically improved even though only the VMC energy is explicitly optimized.

The difference in the behaviors of the C$_2$ and Si$_2$ molecules is striking.
While for Si$_2$ the energy shows a small gradual decrease upon increasing the number of determinants,
for C$_2$ there is a very rapid initial decrease, a manifestation of the presence of
strong energetic near-degeneracies among valence orbitals.
The fixed-node error of a single-determinant wave function using MCSCF natural orbitals is about
$6$ mHa for Si$_2$ and about $46$ mHa for C$_2$, almost an order of magnitude larger!

\begin{figure}
\includegraphics[width=7.6cm]{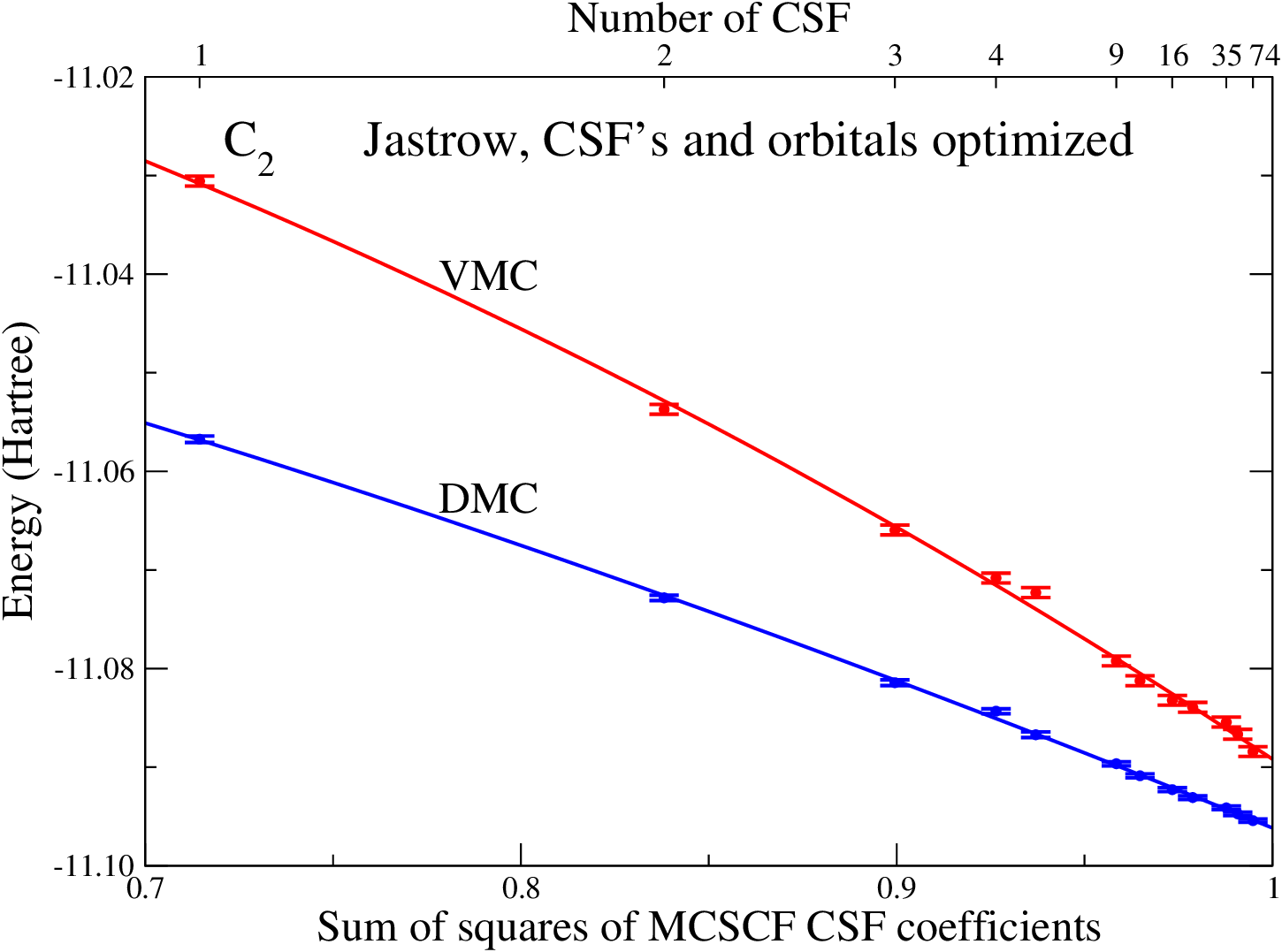}
\caption{(Color online) Extrapolation (to $1$) by quadratic fits of the VMC and DMC energies obtained with the
truncated, fully reoptimized Jastrow-Slater CAS(8,14) wave functions with respect to the sum of the
squares of the MCSCF CSF coefficients retained, $\sum_{i=1}^{\NCSF} (c_i^{\text{MCSCF}})^2$.
}
\label{extrapolation}
\vspace{-0.4cm}
\end{figure}

Retaining all the determinants of a CAS(8,14) wave function would be costly in QMC
but one can estimate the energy in this limit by extrapolation.
Fig.~\ref{extrapolation} shows a quadratic fit of the VMC and DMC energies obtained with truncated,
fully reoptimized % multi-determinantal
wave functions with respect to the sum of the squares of the MCSCF CSF coefficients retained,
$\sum_{i=1}^{\NCSF} (c_i^{\text{MCSCF}})^2$.
This sum is equal to $1$ in the limit that all the CSFs are included.

\begin{table}
\vspace{-0.2cm}
\caption{
Total energy and well depth
of the  C$_2$ molecule, with a HF pseudopotential~\cite{BurkatzkiFilippiDolg06},
obtained in DMC with a series of fully optimized Jastrow-Slater wave functions.
The well depth in column 3 is obtained using a well-converged DMC atomic energy,
-5.43401(6) H, % new psp.
whereas that in column 4 uses atomic CAS wave functions consistent with the molecular
ones in order to achieve some cancellation of error.
}
\begin{tabular}{lccc}
\hline
Wave function & Total energy (H) & \multicolumn{2}{c}{Well depth (eV)}\\
\hline
\hline
1 determinant             & -11.0566(3)& 5.13(1) & 5.70(1) \\
CAS(8,8)                  & -11.0922(3)& 6.10(1) & 6.39(1) \\
CAS(8,10)                 & -11.0939(3)& 6.15(1) & 6.37(1) \\
CAS(8,14)                 & -11.0962(3)& 6.21(1) & 6.38(1) \\
CAS(8,18)                 & -11.0986(3)& 6.28(1) & 6.36(1) \\
RAS(8,26)                 & -11.1007(3)& 6.33(1) & --      \\

Estimated exact$^a$       & --       & 6.36(2) & 6.36(2)\\
\hline
\multicolumn{4}{l}{
$^a$ Scalar-relativistic, valence-corrected estimate of Ref.~\onlinecite{BytRue-JCP-05}.
}
\end{tabular}
\label{cas}
\vspace{-0.4cm}
\end{table}

Table~\ref{cas} reports the extrapolated DMC energies and well depths (dissociation energy + zero-point energy)
for a series of fully optimized Jastrow-Slater CAS and RAS wave functions.
Increasing the size of the active space results in a monotonic improvement of the total energy
and the well depth in column 3 obtained using
a good estimate of the exact atomic energy from a DMC calculation with a CAS(4,13) wave function.
Chemical accuracy (1 kcal/mol = 0.04 eV) is reached for the largest active space.
Alternatively, as shown in column 4, good well depths can be obtained using much smaller active spaces
by relying on a partial cancellation of error
employing atomic wave functions consistent with the molecular ones:
for the molecular single determinant wave function, an atomic single determinant wave function is also used;
for the molecular CAS(8,$m$) wave functions, atomic CAS(4,$m$/2) wave functions are used.
In this case, while the use of a single-determinant wave function yields an error of 0.66 eV,
all the CAS wave functions yield well-depths that agree with the exact one to
better than chemical accuracy.
This behavior parallels the standard quantum chemistry approaches where single-determinant reference methods
such as coupled cluster are in error by as much as 0.2 eV while multi-reference configuration interaction
calculations yield chemical accuracy~\cite{BytRue-JCP-05}.
Density functional theory methods on the other hand perform rather poorly, giving a well depth
of 6.69 and 4.69 eV within LDA and B3LYP, respectively.

\vskip 2mm \noindent {\it Conclusions.} \hskip 2mm
The method presented provides a systematic means of eliminating the main limitation of
present-day QMC calculations, namely the fixed-node error, and supercedes
variance minimization~\cite{UWW88} as the method of choice for optimizing many body wave functions.
Extension of the method to geometry optimization will overcome the other major limitation
of QMC methods.

\vskip 2mm \noindent {\it Acknowledgments.} \hskip 2mm
We thank M. P. Nightingale, R. Assaraf and A. Savin for discussions.
Computations were performed at NERSC, NCSA, OSC and CNF.
Supported by NSF (DMR-0205328, EAR-0530301),
Sandia Natl. Lab.,
FOM (Netherlands) and MIUR-COFIN05.

\bibliographystyle{apsrev}

\bibliography{qmc}

\end{document}